\documentclass{jpsj2}

\usepackage{txfonts}
\usepackage{float}
\usepackage{color}
\usepackage{ulem}

\title{Structural Evolution from Hyper-Honeycomb to Honeycomb Networks and Superconductivity in LaPt$_x$Si$_{2-x}$}

\author{Sitaram Ramakrishnan$^{1}$\thanks{E-mail address: niranj002@gmail.com, Present address: I-HUB Quantum Technology Foundation, Indian Institute of Science Education and Research, Pune, 411008, India.},
Tatsuya Yamakawa$^{1}$,
Ryohei Oishi$^{1}$,
Soichiro Yamane$^{2}$,
Atsutoshi Ikeda$^{2}$,
Masaki Kado$^{1}$,
Yasuyuki Shimura$^{1}$,
Toshiro Takabatake$^{1}$,
Takahiro Onimaru$^{1}$,
Yasuhiro Shibata$^{3}$, 
Arumugam Thamizhavel$^{4}$,
Srinivasan Ramakrishnan$^{5}$,
Shingo Yonezawa$^{2}$,
and Minoru Nohara$^{1}$\thanks{E-mail address: mnohara@hiroshima-u.ac.jp}
}

\inst{
$^{1}$ Department of Quantum Matter, AdSE, Hiroshima University, Higashi-Hiroshima 739-8530, Japan \\
$^{2}$ Department of Electronic Science and Engineering, Graduate School of Engineering, Kyoto University, \\Kyoto 615-8510, Japan\\
$^{3}$ Natural Science Center for Basic Research and Development, Hiroshima University, \\Higashi-Hiroshima 739-8526, Japan\\
$^{4}$Department of Condensed Matter Physics and Materials Science, Tata Institute of Fundamental Research, \\Mumbai 400005, India \\
$^{5}$Department of Physics, Indian Institute of Science Education and Research, Pune, 411008, India  \\
}

\abst{
We report the crystal structures and superconductivity (SC) of LaPt$_{x}$Si$_{2-x}$ ($0.5 \leq x \leq 1.0$) that are solid solutions of LaSi$_{2}$ and LaPtSi with centrosymmetric tetragonal ($I4_{1}/amd$, $D_{4h}^{19}$, \#141) and non-centrosymmetric tetragonal ($I4_{1}md$, $C_{4v}^{11}$, \#109) structures, respectively. It was found that at $0.86 \leq x \leq 1.00$, the non-centrosymmetric tetragonal symmetry is preserved, while partial disorder appears in alternating Pt and Si of the hyper-honeycomb network. The superconducting transition temperature $T_{\rm c}$ was drastically reduced from 3.9 K to 1.5 K as $x$ varies from 1.0 to 0.86. Additionally, a hexagonal phase with an AlB$_{2}$-type structure ($P6/mmm$, $D_{6h}^{1}$, \#191) has been discovered at $0.50 \leq x \leq 0.71$ with a honeycomb network of statistically distributed Pt and Si atoms. The hexagonal phase exhibited SC at $T_{\rm c} = 0.38$ K. This system provides an opportunity to investigate the relationship between topological electronic states, SC, and disorders. 
}

\begin{document}
\maketitle

\section{Introduction}

\begin{figure}
\centering
\includegraphics[width=105mm]{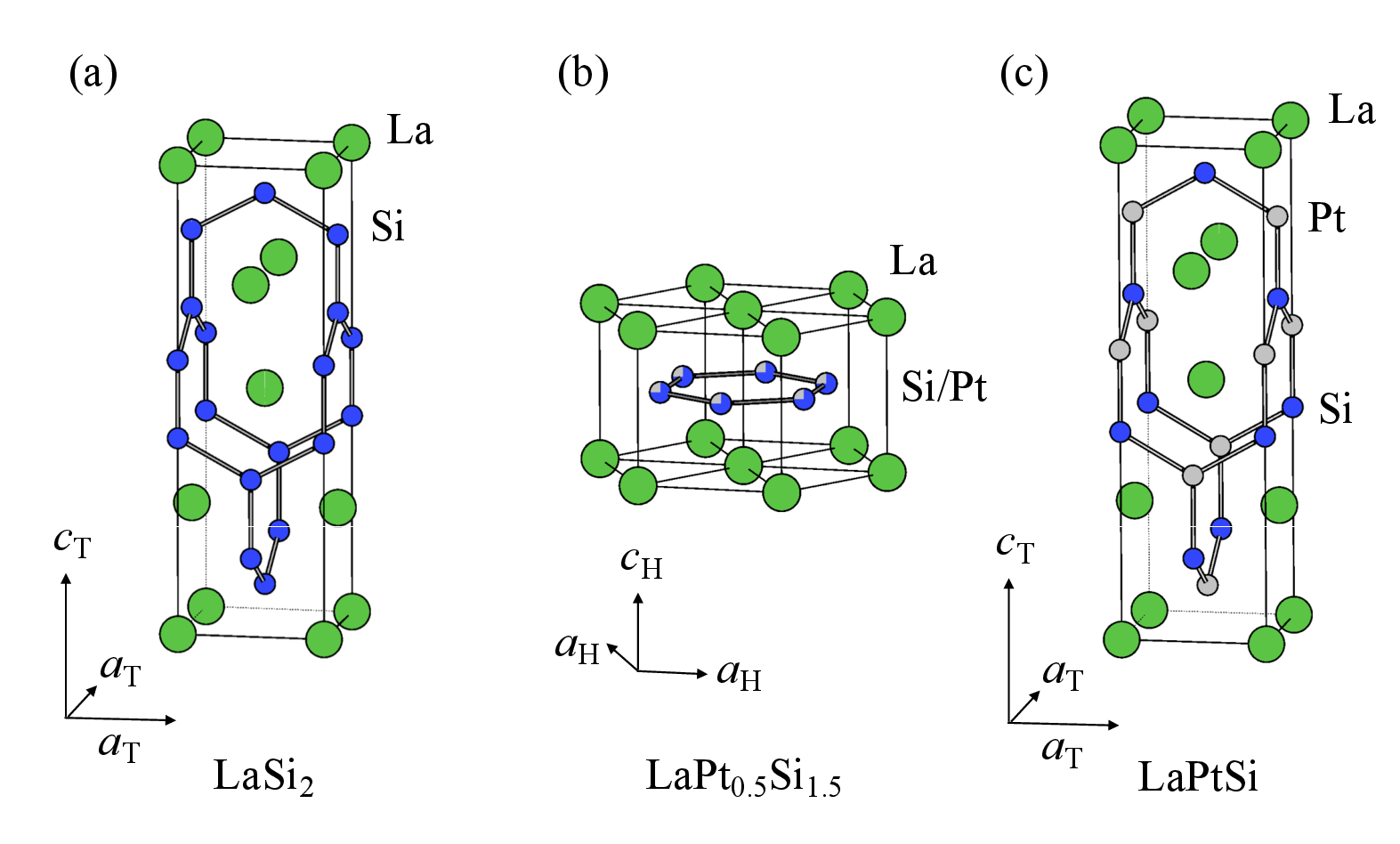}
\caption{(Color online) Crystal structures of three phases of LaPt$_{x}$Si$_{2-x}$.  
 (a) LaSi$_2$ exhibits the tetragonal $\alpha$-ThSi$_2$-type structure ($I4_{1}/amd$, $D_{4h}^{19}$, \#141). 
 (b) LaPt$_{0.5}$Si$_{1.5}$ exhibits the hexagonal AlB$_2$-type structure ($P6/mmm$, $D_{6h}^{1}$, \#191). 
 (c) LaPtSi exhibits the tetragonal LaPtSi-type structure ($I4_{1}md$, $C_{4v}^{11}$, \#109). 
In LaPt$_{0.5}$Si$_{1.5}$, Pt/Si forms a honeycomb network, while in LaSi$_2$ and LaPtSi, it forms a hyper honeycomb network. The structures are centrosymmetric in LaSi$_2$ and LaPt$_{0.5}$Si$_{1.5}$, while non-centrosymmetric in LaPtSi.}
\label{fig:structures}
\end{figure}

Equiatomic ternary silicide LaPtSi and isotypic compounds, such as LaNiSi, LaPtGe, and ThIrSi, are attracting considerable interest because the first-principles calculations suggest topological electronic states and exotic superconductivity (SC) due to their crystal structures with broken spatial inversion symmetry and nonsymmorphic space group. \cite{zhang2020a,shi2021a,ptok2019a}
Additionally, muon-spin rotation and relaxation ($\mu$SR) measurements indicate topological SC with broken time-reversal symmetry (TRS) in LaNiSi, LaPtSi, and LaPtGe, \cite{shang2022a} while the TRS is preserved in ThIrSi.\cite{tay2023a}
These compounds crystallize into the non-centrosymmetric tetragonal structure of LaPtSi-type ($I4_{1}md$, $C_{4v}^{11}$, \#109), which is an ordered variant of the centrosymmetric tetragonal structure of $\alpha$-ThSi$_{2}$-type ($I4_{1}/amd$, $D_{4h}^{19}$, \#141).
In LaSi$_{2}$ with $\alpha$-ThSi$_{2}$-type structure, Si forms a hyper-honeycomb network characterized by three covalent bonds per Si in a twisted manner in a body-centered lattice, as depicted in Fig. \ref{fig:structures}(a).
In contrast, in LaPtSi, the hyper-honeycomb network is formed by alternating Pt and Si atoms, thereby breaking the center of inversion, as depicted in Fig. \ref{fig:structures}(c), originally reported by Klepp and Parth$\rm{\acute{e}}$ using powder X-ray diffraction (PXRD). \cite{klepp1982a} 
Note that $I4_{1}md$ is a subgroup of $I4_{1}/amd$.

SC in the $\alpha$-ThSi$_{2}$-type structure has been observed in the centrosymmetric compounds such as LaSi$_2$ with a superconducting transition temperature $T_{\rm c}$ = 2.3 K. \cite{hardy1954a,matthias1958a,Ohtsuka1966a,satoh1966a,mcwhan1967a,satoh1970a,evers1980a,iyo2017a}
For the non-centrosymmetric counterpart, SC in LaPtSi was discovered at $T_{\rm c}$ = 3.3 K by Evers et al. through AC susceptibility measurements. \cite{evers1984a} 
Ramakrishnan et al. \cite{ramakrishnan1995a} revealed that LaPtSi is a full-gap and weak-coupling $s$-wave superconductor in the dirty limit, where the $T_{\rm c}$ was comparatively higher at 3.9 K than the value reported by the earlier study. \cite{evers1984a} Later, Kneidinger et al. \cite{kneidinger2013a} observed similar behavior in the SC of LaPtSi with a $T_{\rm c}$ = 3.35 K, close to that of Evers et al. \cite{evers1984a}
Furthermore, penetration depth measurements indicated dirty and full-gap $s$-wave SC in LaPtSi. \cite{palazzese2018a}
These observations suggest atomic disorders, which may impact possible topological band structures and SC in LaPtSi. 

Atomic disorders are observed in $\alpha$-ThSi$_{2}$-type solid solutions, ThRh$_{x}$Si$_{2-x}$ and ThIr$_{x}$Si$_{2-x}$, which exhibit a large homogeneity range of $0 \leq x \leq 0.96$ and $0 \leq x \leq 1.0$, respectively. \cite{lejay1983a,chevalier1986a}
Interestingly, $T_{\rm c}$ decreases from 3.2 K at $x$ = 0.0 to below 1.7 K at $x \simeq$ 0.25, but reappears at $x > 0.7$ and reaches a maximum of 6.5 K at $x$ = 0.96 and 1.0, respectively. \cite{lejay1983a,chevalier1986a}
PXRD inferred that Rh/Ir and Si were distributed statistically over the hyper-honeycomb network, making the system the $\alpha$-ThSi$_{2}$-type structure. \cite{lejay1983a,chevalier1986a}
However, Klepp and Parth$\rm{\acute{e}}$ discussed the difficulty in determining atomic disorder from PXRD in this class of compounds. \cite{klepp1982a} 
Braun mentioned that the $T_{\rm c}$ variation at $x > 0.7$ is rather suggestive of an atomic ordering phenomenon. \cite{braun1984a}
In other words, the LaPtSi-type structure, rather than the $\alpha$-ThSi$_{2}$-type, is suggested at $x > 0.7$, although a question remains on the crystal structure transition from the $\alpha$-ThSi$_{2}$-type to LaPtSi-type with $x$ in ThRh$_{x}$Si$_{2-x}$ and ThIr$_{x}$Si$_{2-x}$. 
It is interesting to note that for ThCo$_{x}$Si$_{2-x}$ and ThNi$_{x}$Si$_{2-x}$, the hexagonal AlB$_{2}$-type structure ($P6/mmm$, $D_{6h}^{1}$, \#191) emerges at $x \simeq 0.5$ in between the tetragonal $\alpha$-ThSi$_{2}$-type phases at the smaller and larger $x$ with miscibility gaps (phase separations) between them. \cite{zhong1985a,albering1994a}

In this paper, we report the assessment of crystal structures and SC in LaPt$_{x}$Si$_{2-x}$ ($0.5 \leq x \leq 1.0$), solid solutions of LaSi$_{2}$ and LaPtSi with $\alpha$-ThSi$_{2}$-type and LaPtSi-type structures, respectively. 
We grew a single crystal of LaPtSi by the Czochralski (Cz) method resulting in a crystal with composition LaPt$_{0.88}$Si$_{1.12}$ and prepared polycrystalline samples of LaPt$_{x}$Si$_{2-x}$ by arc melting with $x$ = 0.50, 0.75, 0.80, 0.85, and 1.00. 
It is observed that the tetragonal LaPtSi-type structure remains undisturbed for the range of $0.86 \leq x \leq 1.00$ in LaPt$_{x}$Si$_{2-x}$, while the inter-site mixing of Pt and Si exists within the hyper-honeycomb network, confirmed by the refinement of single-crystal X-ray diffraction (SXRD) data on LaPt$_{0.88}$Si$_{1.12}$. 
The system exhibits a significant decrease in $T_{\rm c}$ from 3.9 K \cite{ramakrishnan1995a} at $x$ = 1.0 to 1.5 K at $x$ = 0.86, suggesting the effects of disorder on SC.  
By further reducing the Pt content $x$, a phase separation occurs, ultimately leading to the formation of a hexagonal AlB$_2$-type phase ($P6/mmm$, $D_{6h}^{1}$, \#191) in LaPt$_{x}$Si$_{2-x}$ within the range of $0.50 \leq x \leq 0.71$. 
We have newly discovered that this hexagonal phase is also superconducting, albeit with a reduced $T_{\rm c}$ of 0.38 K at $x$ = 0.50. 

\section{Experimental}

A single crystal of LaPt$_{0.88}$Si$_{1.12}$ was grown using a modified Cz method. Polycrystalline samples of LaPt$_{x}$Si$_{2-x}$ with nominal compositions of $x$ = 0.50, 0.75, 0.80, 0.85, and 1.00 were synthesized by arc-melting. The composition of the samples was determined using an electron-probe microanalyzer (EPMA) (JXA-iSP100) with an accelerating voltage of 20 kV and a beam current of 30 nA. Hereafter, we denote nominal and analyzed values by $x_{\rm nominal}$ and $x_{\rm analyzed}$, respectively.
Further details of sample preparation are described in the Appendix\ref{app:synthesis}.

SXRD was measured on a four-circle Bruker diffractometer employing Mo K$\alpha$ radiation for LaPt$_{0.88}$Si$_{1.12}$.
SXRD data were processed by the APEX-III software. \cite{apex3} 
Structure refinements were done using JANA 2006. \cite{petricekv2014a} 
The results from SXRD are consistent with the composition determined by EPMA, resulting in LaPt$_{0.88}$Si$_{1.12}$.
The crystallographic table of SXRD is given in the Appendix\ref{app:sxrd}.
PXRD was measured on a Rigaku MiniFlex600 with Cu K$\alpha$ radiation and the Rietveld refinement was done for LaPt$_{x}$Si$_{2-x}$ with $x$ = 0.50 using FULLPROF software package. \cite{juan1993a}

For single-crystalline LaPt$_{0.88}$Si$_{1.12}$,
the electrical resistivity was measured by the standard DC four probe method in a Physical Property Measurement System (PPMS, Quantum Design, USA). 
The specific heat measurements were carried out in the PPMS with a $^3$He refrigerator. 
A superconducting quantum interference device (SQUID) magnetometer (MPMS, Quantum Design, USA)
was used to measure magnetization $M$.
For polycrystalline LaPt$_{x}$Si$_{2-x}$, 
the electrical resistivity was measured using the PPMS with an adiabatic demagnetization refrigerator (ADR). 
AC susceptibility measurements were carried out using a home-built susceptometer, which fits in the PPMS with ADR. Details regarding the instrumentation is described by Yonezawa et al. \cite{yonezawa2015a}

\section{Results and discussion}

\subsection{\label{subec:Laptsi_XRD} Tetragonal and hexagonal phases in LaPt$_{x}$Si$_{2-x}$}

Table \ref{epma} shows the analyzed composition for LaPt$_{x}$Si$_{2-x}$, where one observes that 
for $x_{\rm nominal}$ = 1.00 and 0.50, the $x_{\rm analyzed}$ is relatively close to $x_{\rm nominal}$. 
However, for $x_{\rm nominal}$ = 0.75, 0.80, and 0.85, 
there is a phase separation resulting in both hexagonal AlB$_2$-type and tetragonal LaPtSi-type phases, 
as confirmed by PXRD. Details of EPMA and PXRD are described in the Appendix\ref{app:epma} and\ref{app:pxrd}, respectively.

\begin{table}
\centering
\small
\caption{
Chemical composition and crystal structure of LaPt$_{x}$Si$_{2-x}$. 
$x_{\rm nominal}$ represents the prescribed value of $x$, while $x_{\rm analyzed}$ is determined by EPMA. 
T and H denote the tetragonal LaPtSi-type ($I4_{1}md$, $C_{4v}^{11}$, \#109)
and the hexagonal AlB$_2$-type ($P6/mmm$, $D_{6h}^{1}$, \#191) structures, respectively.}
\label{epma}%
\begin{tabular}{ccc}
\hline
$x_{\rm nominal}$ & $x_{\rm analyzed}$ (Structure) &  Type \\
\hline
1.00      &    0.98(2) (T)  & Polycrystal \\
1.00     &   0.88(3) (T)  & Single crystal \\
0.85     & 0.708(5) (H) + 0.86(4) (T) & Polycrystal \\
0.80     &  0.692(4) (H) + 0.869(5) (T) & Polycrystal \\
0.75     &  0.691(4) (H) + 0.870(7) (T) & Polycrystal \\
0.50     &   0.493(9) (H)  & Polycrystal  \\
\hline
\end{tabular}
\end{table}

For single-crystalline LaPt$_{0.88}$Si$_{1.12}$ in the tetragonal phase,
all diffraction maxima were indexed by a single unit cell with lattice parameters
$a_{\rm T}$ = 4.2441(2) {\AA} and $c_{\rm T}$ = 14.5264(2) {\AA}, similar to the published unit
cell of LaPtSi \cite{klepp1982a, ramakrishnan1995a, kneidinger2013a, palazzese2018a},
albeit with a slight reduction in volume. 
Table \ref{atomtetra} shows the atomic coordinates for tetragonal LaPt$_{0.88}$Si$_{1.12}$. 
Figure \ref{fig:structures}(c) shows the crystal structure of LaPtSi, which is isostructural to LaPt$_{0.88}$Si$_{1.12}$.  
They are derivatives of the $\alpha$-ThSi$_2$-type structure, like LaSi$_2$ \cite{satoh1970a} or SrGe$_2$ \cite{akira2017a}, where the Si/Ge atoms form a hyper-honeycomb network in a body-centered lattice, resulting in a centrosymmetric ($I4{_1}/{amd}$) structure, as shown in Fig. \ref{fig:structures}(a). 
In LaPtSi, 50\% of the Si atoms are replaced by Pt, and the hyper-honeycomb network is formed by alternating Pt and Si atoms, thereby breaking the center of inversion and crystallizing into the non-centrosymmetric structure ($I4{_1}md$), which is a subgroup of $I4{_1}/{amd}$. 
For LaPt$_{0.88}$Si$_{1.12}$, the hyper-honeycomb network is formed by alternating Pt-rich and Si-rich sites, preserving $I4{_1}md$ symmetry, thereby the structure is the LaPtSi-type.

\begin{table*}
\centering
\small
\caption{
Structural parameters for the tetragonal structure of crystal of LaPt$_{0.88}$Si$_{1.12}$  at 300 K.
Given are the occupancy, the fractional coordinates $x$, $y$, $z$ of the atoms,
their anisotropic displacement parameters (ADPs) $U_{ij}$ $(i, j = 1, 2, 3)$, 
and the equivalent isotropic displacement parameter $U^{\rm eq}_{\rm iso}$.
The off-diagonal elements $U_{12}$, $U_{13}$, and $U_{23}$ are zero because all atoms in the crystal structure occupy the $4a$ Wyckoff positions.
$R_F$(obs/all) = 0.0311/0.0363, no. of parameters is 15.
Refinement method used: Least-squares on $F$. 
Space group: $I4_{1}md$, $C_{4v}^{11}$, \# 109.
Criterion of observability: $I < 3\sigma(I)$.}
\label{atomtetra}%
\begin{tabular}{cccccccccccc}
\hline
Atom & Occupancy  & $x$ & $y$ & $z$ & $U_{11}$ & $U_{22}$ & $U_{33}$ & $U^{\rm eq}_{\rm iso}$ \\
\hline
La &1 &0 & 0 & 0 &  0.0086(18) &   0.0050(19) &  0.0098(19) &  0.0078(11) \\
Pt &0.83(3) &0  &0 &0.5852(2) &  0.0108(10) &   0.0010(10) &  0.0042(9) &  0.0053(6) \\
Si$_{\rm Pt}$ &0.17 &0  &0 &0.5852 &  0.0108 &   0.0010 &  0.0042 &  0.0053 \\
Si &0.95(3) &0 &0 &0.4188(7) &   0.014(6) & 0.009(7) &  0.008(6)  &   0.011(4) \\
Pt$_{\rm Si}$ & 0.05 &0 &0 &0.4188 &   0.014 & 0.009 &  0.008 &   0.011 \\
\hline
\end{tabular}
\end{table*}

\begin{figure}
\centering
\includegraphics[width=90mm]{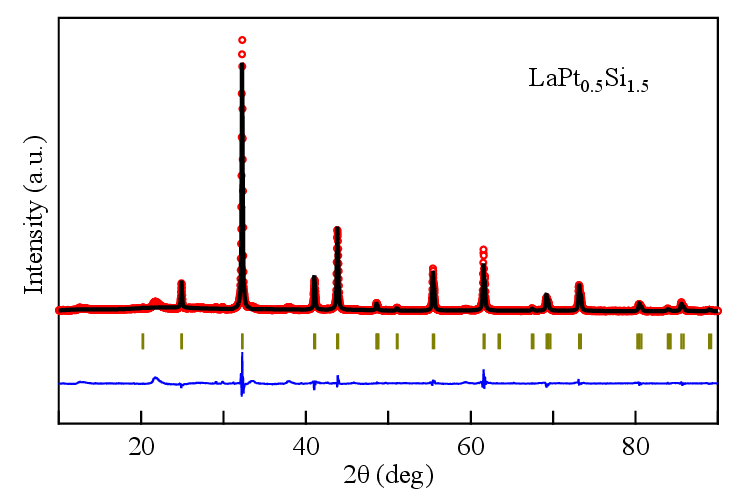}
 \caption{(Color online) 
The PXRD pattern of LaPt$_{0.5}$Si$_{1.5}$ at room temperature.
Red open circles, black and blue lines, and green bars indicate observed diffraction, fit, residual, and Bragg peak positions, respectively. 
The Rietveld refinement revealed a hexagonal structure, 
space group $P6/mmm$ ($D_{6h}^{1}$, \#191), 
lattice parameters $a_{\rm H}$ = 4.13149(5) {\AA} and $c_{\rm H}$ = 4.39936(7) {\AA}, 
and atomic positions of 
La at $1a$ (0, 0, 0), occupancy 1; 
Pt at $2d$ (1/3 , 2/3, 1/2), occupancy 0.25;  
Si at $2d$ (1/3 , 2/3, 1/2), occupancy 0.75. 
A good fit is obtained with Bragg $R_F$ = 5.55\% and $\chi^{2}$ = 5.2101.
 }
\label{fig:pxrd_hex}
\end{figure}

With further reduction of $x$ down to 0.5, we have discovered a hexagonal phase as indicated in Fig. \ref{fig:structures}(b). 
Figure \ref{fig:pxrd_hex} shows the PXRD pattern along with the Rietveld refinement for the hexagonal phase
with lattice parameters $a_{\rm H}$ = 4.13149(5) {\AA} and $c_{\rm H}$ = 4.39936(7) {\AA} with the space group $P6/mmm$, 
revealing that LaPt$_{0.5}$Si$_{1.5}$ crystallizes into the hexagonal AlB$_{2}$-type structure. 

\begin{figure}
\centering
\includegraphics[width=90mm]{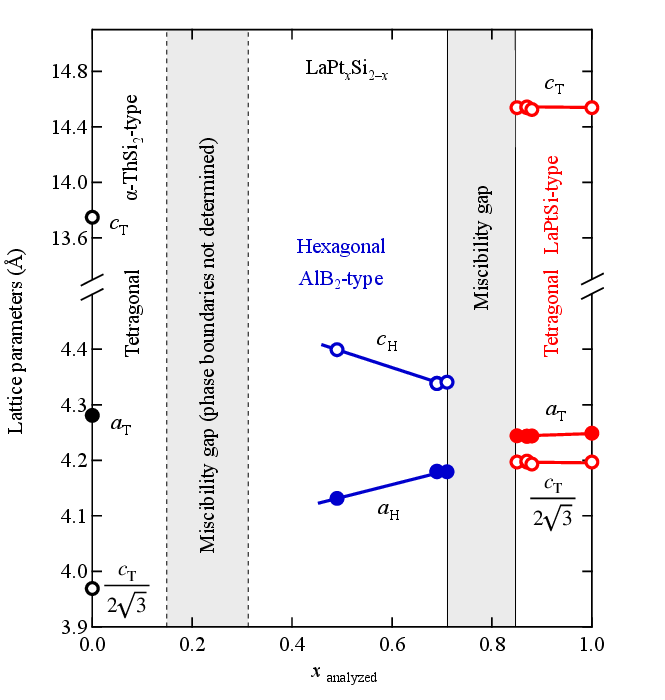}
 \caption{(Color online) Structural phase diagram of LaPt$_{x}$Si$_{2-x}$. 
 Lattice parameters are plotted as a function of $x_{\rm analyzed}$, which was determined by EPMA. 
 $a_{\rm T}$ and $c_{\rm T}$ denote lattice parameters of the tetragonal structures, while
 $a_{\rm H}$ and $c_{\rm H}$ denote lattice parameters of the hexagonal structure. 
 $c_{\rm T}/2\sqrt{3}$ is plotted to compare with $a_{\rm T}$ to characterize the hyper-honeycomb network. 
At $x \geq 0.86$, the crystal structure is tetragonal LaPtSi-type ($I4_{1}md$, $C_{4v}^{11}$, \#109). 
A separation of two phases is observed for $0.71 < x < 0.86$. 
Within the range of $0.50 \leq x \leq 0.71$, the crystal structure is hexagonal AlB$_2$-type ($P6/mmm$, $D_{6h}^{1}$, \#191).
Currently, it is not known what happens at $0 < x < 0.50$,
perhaps there is yet another phase separation. 
At $x = 0.0$, the crystal structure is tetragonal $\alpha$-ThSi$_2$-type structure ($I4_{1}/amd$, $D_{4h}^{19}$, \#141).
}
\label{fig:lattice_parameters}
\end{figure}

The change from tetragonal to hexagonal structure is further illustrated by a phase diagram depicting the variation in lattice parameters, as shown in Fig. \ref{fig:lattice_parameters}. 
It is obvious from Fig. \ref{fig:lattice_parameters}, for $x \geq 0.86$ the system remains tetragonal LaPtSi-type phase.
However, for $x$ in the range of $0.71 < x < 0.86$, a clear phase separation has been observed through EPMA. 
PXRD patterns, EPMA images and a table describing the phase separation in detail are given in the Appendix\ref{app:epma} and\ref{app:pxrd}. 
Currently it is unclear what happens in the range $x < 0.5$. 
However, on comparison with a similar compound CePt$_x$Si$_{2-x}$ \cite{gribanov2004a},
a phase separation between tetragonal $\alpha$-ThSi$_{2}$-type and hexagonal AlB$_{2}$-type phases can be expected. 

A honeycomb network is characterized by a lattice parameter $a_{\rm H}$ of the hexagonal structure. 
The corresponding length in a hyper-honeycomb network can be expressed by two parameters $a_{\rm T}$ and $c_{\rm T}/2\sqrt{3}$ using tetragonal lattice parameters. 
In the LaPtSi phase, $a_{\rm T}$ is similar to $c_{\rm T}/2\sqrt{3}$, while in the LaSi$_{2}$, $a_{\rm T}$ is significantly larger than $c_{\rm T}/2\sqrt{3}$, as shown in Fig. \ref{fig:lattice_parameters}, suggesting the distortion of the hyper-honeycomb network is small in the LaPtSi phase. In addition, $a_{\rm H}$ at the phase boundary of the hexagonal AlB$_{2}$-type phase is close to $a_{\rm T}$ and $c_{\rm T}/2\sqrt{3}$ of the tetragonal LaPtSi-type phase. 
This suggests that despite there being no group-subgroup relation between $P6/mmm$ and $I4{_1}{md}$, the characteristic feature of the body-centered lattice in the tetragonal phase which is the hyper-honeycomb network of Pt/Si atoms allows a distortion toward a honeycomb network, thereby resembling the AlB$_2$-type structure.

\subsection{\label{Superconductivity}Phase diagram of superconductivity in LaPt$_{x}$Si$_{2-x}$}

\begin{figure}
\centering
\includegraphics[width=90mm]{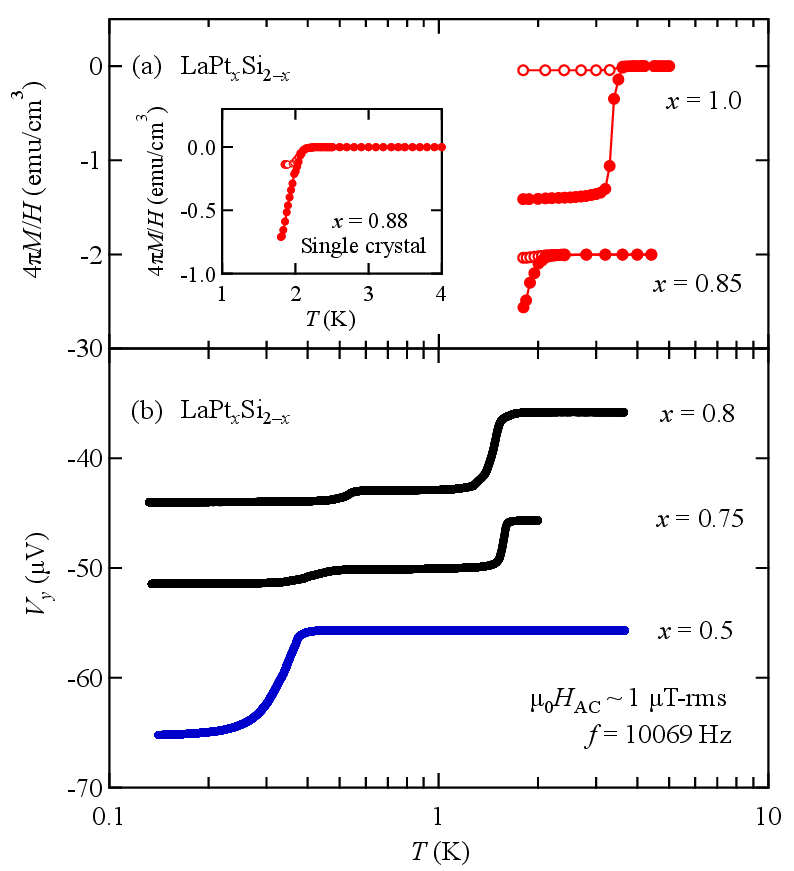}
 \caption{(Color online) 
(a) Temperature dependence of the magnetization divided by field $M/H$ of LaPt$_{x}$Si$_{2-x}$ with $x$ = 1.0 and 0.85 in a field of 1 mT in the zero-field-cooled (ZFC) and field-cooled (FC) conditions. The inset shows $M/H$ of single-crystalline LaPt$_{x}$Si$_{2-x}$ with $x$ = 0.88.  
(b) Real part of AC magnetic susceptibility $V_{y}$ of LaPt$_{x}$Si$_{2-x}$ with $x$ = 0.8, 0.75, and 0.50. 
The data are shifted vertically for the sake of clarity. 
}
\label{fig:chi}
\end{figure}

\begin{figure}
\centering
\includegraphics[width=90mm]{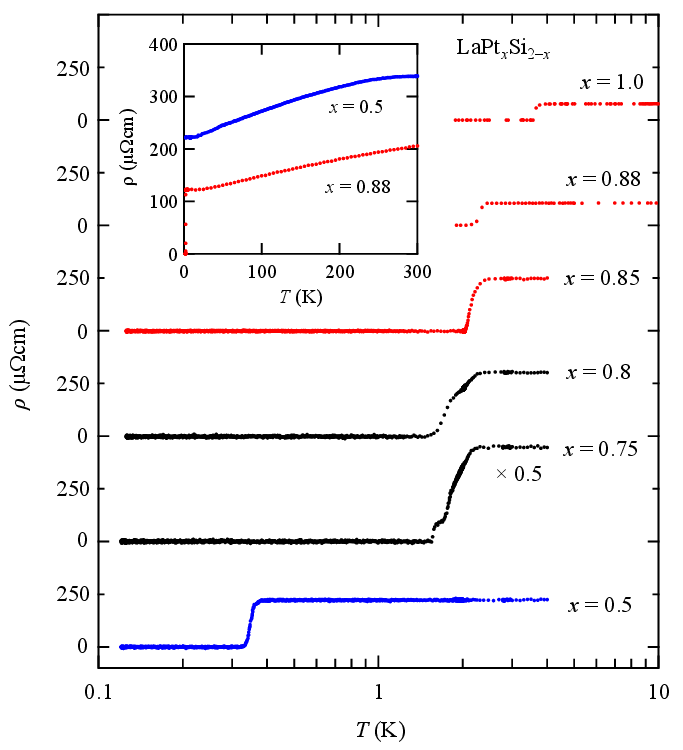}
 \caption{(Color online) 
Temperature dependence of the electrical resistivity $\rho$ for polycrystalline LaPt$_{x}$Si$_{2-x}$ at nominal values of $x$ = 1.0, 0.85, 0.80, 0.75, and 0.50, as well as for a single crystal of LaPt$_{x}$Si$_{2-x}$ with an analyzed value of $x$ = 0.88. 
The inset shows $\rho$ for $x$ = 0.88 (single crystal) and 0.50 (poly crystal) over a wide temperature range. 
}
\label{fig:r_SC}
\end{figure}

Figure \ref{fig:chi} shows the results of magnetic measurements in LaPt$_{x}$Si$_{2-x}$. SC is observed at $T_{\rm c} = 3.6$ and 2.2 K for $x = 1.00$ and 0.88, respectively, in the tetragonal LaPtSi-type phase, while $T_{\rm c} = 0.38$ K for $x = 0.50$ in the hexagonal AlB$_{2}$-type phase. For $x = 0.80$ and 0.75, two superconducting transitions are observed due to phase separation. SC with lower $T_{\rm c}$ is attributed to the hexagonal phase, while that with higher $T_{\rm c}$ is attributed to the tetragonal phase, based on changes in the real part of AC susceptibility $V_{y}$ at $T_{\rm c}$ with respect to $x$. 
Figure \ref{fig:r_SC} shows electrical resistivity for LaPt$_{x}$Si$_{2-x}$. Zero resistivity was observed at $T_{\rm c} = 3.56$ and 2.05 K for $x = 1.0$ and 0.88 in the tetragonal phase, respectively, while $T_{\rm c} = 0.33$ K for $x = 0.50$ in the hexagonal phase. Resistivity exhibited the higher-$T_{\rm c}$ transitions for the phase separated samples ($x = 0.75$, 0.80, and 0.85).
For $x$ = 0.50, a superconducting volume of 60--70\% is estimated from the change $\Delta V_{y}$ below $T_{\rm c}$ by comparing the superconducting transition of the Al standard. \cite{yonezawa2015a}

\begin{figure}
\centering
\includegraphics[width=100mm]{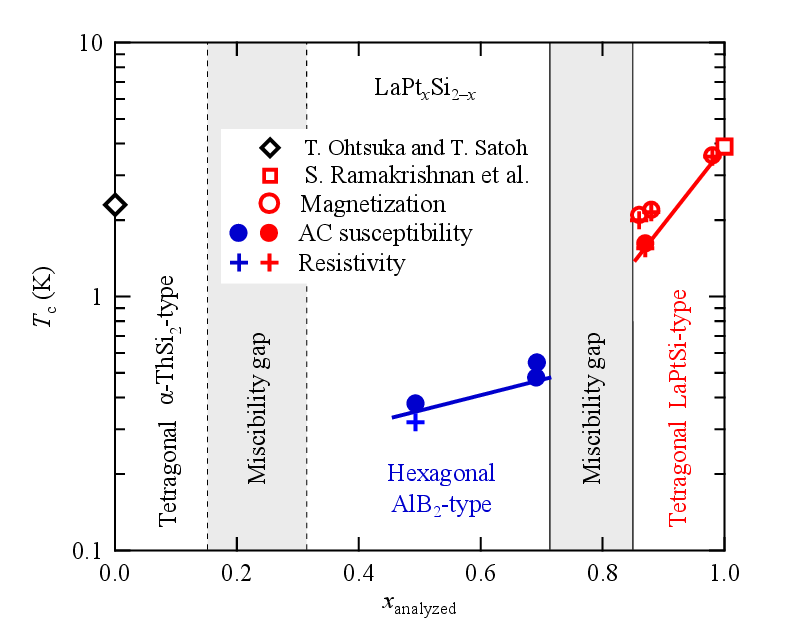}
 \caption{(Color online) Relationship between the analyzed value of $x$
and the superconducting transition temperature $T_{\rm c}$ in LaPt$_{x}$Si$_{2-x}$.
The $T_{\rm c}$ values for LaSi$_2$ ($x = 0.0$) and LaPtSi ($x = 1.0$) are taken from T. Ohtsuka and T. Satoh, \cite{Ohtsuka1966a} and S. Ramakrishnan et al.,\cite{ramakrishnan1995a} respectively, 
and were determined from specific heat measurements.}
\label{fig:Tc_vs_x}
\end{figure}

SC in hexagonal and tetragonal phases is further elucidated by a phase diagram depicting the change in $T_{\rm c}$ as a function of $x_{\rm analyzed}$. From Fig. \ref{fig:Tc_vs_x}, it is evident that in the tetragonal phase, $T_{\rm c}$ decreases markedly with decreasing $x_{\rm analyzed}$: 
LaPt$_{0.88}$Si$_{1.12}$ shows SC at 2.05 K, which is lower than the values of 3.9 K and 3.35 K reported for LaPtSi.  \cite{ramakrishnan1995a, kneidinger2013a}
The smaller value of residual resistivity ratio (RRR) of 1.5 (as shown in the inset of Fig. \ref{fig:r_SC}), compared to earlier works \cite{ramakrishnan1995a, kneidinger2013a}, suggests disorder in the present sample, which can be a possible origin of the reduced $T_{\rm c}$. 
In contrast, the electronic specific-heat coefficient $\gamma = 8.11$ mJ/K$^{2}$mol for the present LaPt$_{0.88}$Si$_{1.12}$ (as described later) is considerably larger than 3.56 mJ/K$^{2}$mol and 6.5 mJ/K$^{2}$mol  for LaPtSi.  \cite{ramakrishnan1995a, kneidinger2013a}
In the hexagonal phase, $T_{\rm c}$ depends weakly on $x_{\rm analyzed}$.

\subsection{Upper critical fields in LaPt$_{0.5}$Si$_{1.5}$ and LaPt$_{0.88}$Si$_{1.12}$}

\begin{figure}
\centering
\includegraphics[width=90mm]{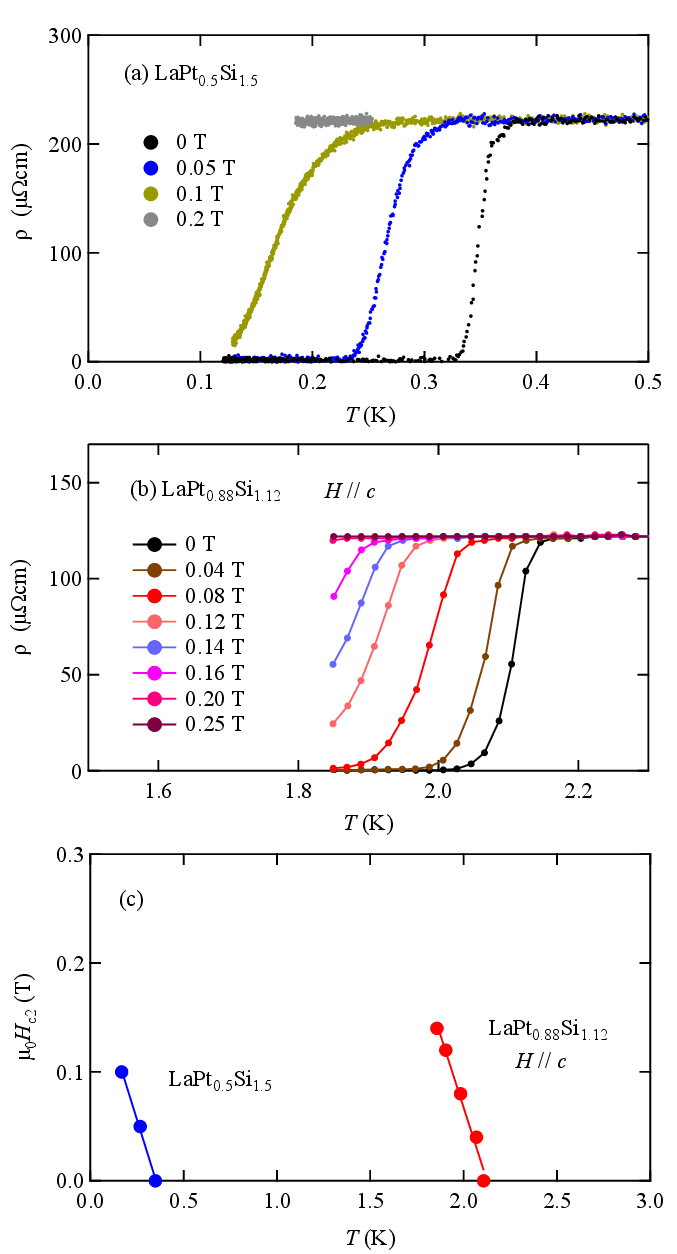}
 \caption{(Color online) 
 (a) The temperature dependence of the electrical resistivity of LaPt$_{0.5}$Si$_{1.5}$ in the hexagonal phase under various magnetic fields. (b) The temperature dependence of the electrical resistivity of single-crystalline LaPt$_{0.88}$Si$_{1.12}$ in the tetragonal phase under various magnetic fields. 
(c) The temperature dependence of the upper critical field $H_{\rm c2}$ of LaPt$_{0.5}$Si$_{1.5}$ and LaPt$_{0.88}$Si$_{1.12}$. The straight lines represent linear fits, from which we estimated $\mu_{0}H_{\rm c2}(0)$ to be 0.19 and 1.1 T for LaPt$_{0.5}$Si$_{1.5}$ and LaPt$_{0.88}$Si$_{1.12}$, respectively. }
\label{fig:Hc2}
\end{figure}

The superconducting transition temperature $T_{\rm c}$ decreases with magnetic field $H$ in the hexagonal phase of LaPt$_{0.5}$Si$_{1.5}$ and the tetragonal phase of LaPt$_{0.88}$Si$_{1.12}$, as shown in Figs. \ref{fig:Hc2}(a) and \ref{fig:Hc2}(b), respectively. 
We defined the transition temperature $T_{\rm c}(H)$ at the temperature where resistivity reaches 50\% of the residual resistivity, and the upper critical field $H_{\rm c2}(T)$ was plotted against temperature $T$ in Fig. \ref{fig:Hc2}(c). 
Linear fits gave estimates of $\mu_{0} H_{\rm c2}(0)$ = 0.19 and 1.1 T for LaPt$_{0.5}$Si$_{1.5}$ and LaPt$_{0.88}$Si$_{1.12}$, respectively. 

Here, we compare $H_{\rm c2}(0)$ with the Pauli limit $H_{\rm P}$, which is given by $\mu_{0} H_{\rm P} \simeq 1.86T_{\rm c}$. For $T_{\rm c}$ = 0.33 K and 2.2 K, the corresponding $\mu_{0}H_{\rm P}$ values are 0.61 T and 4.1 T for LaPt$_{0.5}$Si$_{1.5}$ and LaPt$_{0.88}$Si$_{1.12}$, respectively, which are significantly higher than the observed $H_{\rm c2}(0)$. This indicates that the upper critical fields in these compounds are governed by the orbital limit. LaPtSi has also been reported to be in the orbital limit. \cite{ramakrishnan1995a,kneidinger2013a}
In this context, our data suggest conventional $s$-wave superconductivity, indicating that exotic superconducting states, such as mixed spin-singlet and spin-triplet pairing, are not present in the non-centrosymmetric LaPt$_{0.88}$Si$_{1.12}$. 

We emphasize the importance of investigating whether topological electronic states, such as Weyl nodal rings around $X$ points \cite{zhang2020a} and bulk Dirac points \cite{shi2021a}, remain stable against site mixing, which is inevitable in LaPt$_{x}$Si$_{2-x}$, as demonstrated in the present study. 
Such an investigation is crucial for realizing topological superconductivity, which can emerge from the combination of topological electronic states and $s$-wave superconductivity. \cite{shi2021a}

\subsection{\label{SH}Specific heat in LaPt$_{0.5}$Si$_{1.5}$ and LaPt$_{0.88}$Si$_{1.12}$}

\begin{figure}
\centering
\includegraphics[width=90mm]{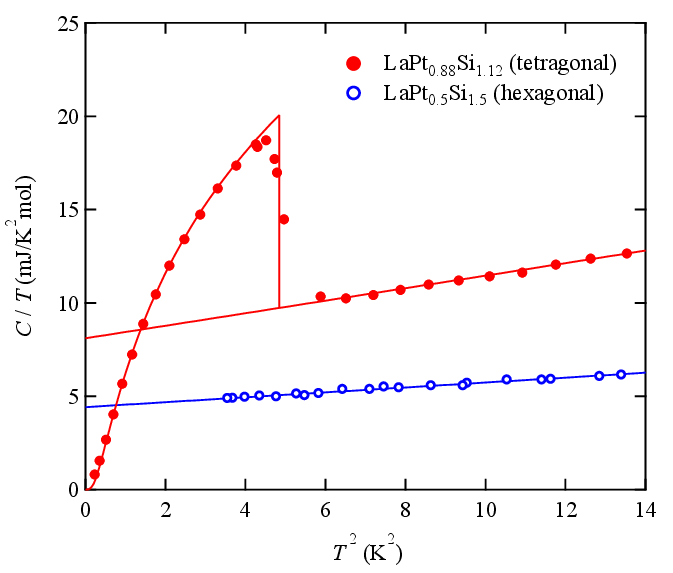}
 \caption{
(Color online) Specific heat divided by temperature, $C/T$, as a function of $T^{2}$ for 
LaPt$_{0.88}$Si$_{1.12}$ (tetragonal phase) and
LaPt$_{0.5}$Si$_{1.5}$ (hexagonal phase). 
The straight lines represent fits by $C/T = \gamma + \beta T^{2}$, where $\gamma$ is the electronic specific heat coefficient and $\beta$ is a constant corresponding to the Debye phonon contributions.
}
\label{fig:Cp}
\end{figure}

Figure \ref{fig:Cp} illustrates the specific heat divided by the temperature $C/T$ as a function of the squared temperature $T^{2}$ for LaPt$_{0.5}$Si$_{1.5}$ in the hexagonal phase and LaPt$_{0.88}$Si$_{1.12}$ in the tetragonal phase.  
Normal-state data above $T_{\rm c}$ can be well-fitted by $C/T = \gamma + \beta T^{2}$, 
where $\gamma$ represents the electronic specific heat coefficient and $\beta$ represents the phonon contributions, from which the Debye temperature $\Theta_{\rm D}$ is estimated. 
In the tetragonal phase of LaPt$_{0.88}$Si$_{1.12}$, a notable jump in specific heat occurs around 2.1 K, which corroborates the $T_{\rm c}$ determined from both magnetization and resistivity measurements. 
The ratio of this jump $\Delta C/\gamma T_{\rm c}$ is calculated as 1.23, with $\gamma$ being 8.11 mJ/K$^2$mol, comparable to the value of the BCS weak coupling limit (1.43), indicating the bulk nature of SC.
The Debye temperature $\Theta_{\rm D}$ is estimated as 259 K from $\beta$ = 0.335 mJ/K$^4$mol, 
using the equation $\Theta_{\rm D} = (12NR\pi^{4}/5\beta)^{1/3}$, 
where $N$ represents the number of atoms in the formula and $R$ denotes the gas constant.
In the hexagonal phase of LaPt$_{0.5}$Si$_{1.5}$, 
$\gamma$ is estimated as 4.42 mJ/K$^2$mol,  $\beta$ as 0.132 mJ/K$^4$mol, and $\Theta_{\rm D}$ = 353 K.
Since SC occurs at a much lower temperature of 0.38 K for this phase, our experimental setup did not allow us to observe the jump at even lower temperatures. However, the AC susceptibility indicates the bulk nature of SC. 

One can estimate thermodynamic critical field $H_{\rm c}(0)$ at $T$ = 0 K using $\gamma$ and $T_{\rm c}$ values from the ratio of $H_{\rm c}(0)^{2}/\gamma T_{\rm c}^2$ = 5.94 for the BCS weak coupling limit. \cite{Padamsee}
Using the aforementioned values of $\gamma$ and $T_{\rm c}$, 
we estimated $\mu_{0}H_{\rm c}(0)$ = 2.87 and 23.2 mT for LaPt$_{0.5}$Si$_{1.5}$ and LaPt$_{0.88}$Si$_{1.12}$, respectively. 
The Ginzburg-Landau parameter was estimated to be $\kappa = H_{\rm c2}(0) / \sqrt{2}H_{\rm c}(0)$ = 47 and 33 for  LaPt$_{0.5}$Si$_{1.5}$ and LaPt$_{0.88}$Si$_{1.12}$, respectively, which were larger than $1/\sqrt{2}$, indicating the type-II SC of both compounds. 
The Ginzburg-Landau coherence length was estimated to be $\xi_{0}$ = 420 and 170 {\AA} for LaPt$_{0.5}$Si$_{1.5}$ and LaPt$_{0.88}$Si$_{1.12}$, respectively, using $\mu_{0}H_{\rm c}(0) = \Phi_{0}/2\pi\xi_{0}^{2}$, where $\Phi_{0}$ is the magnetic flux quantum.

\begin{figure}
\centering
\includegraphics[width=90mm]{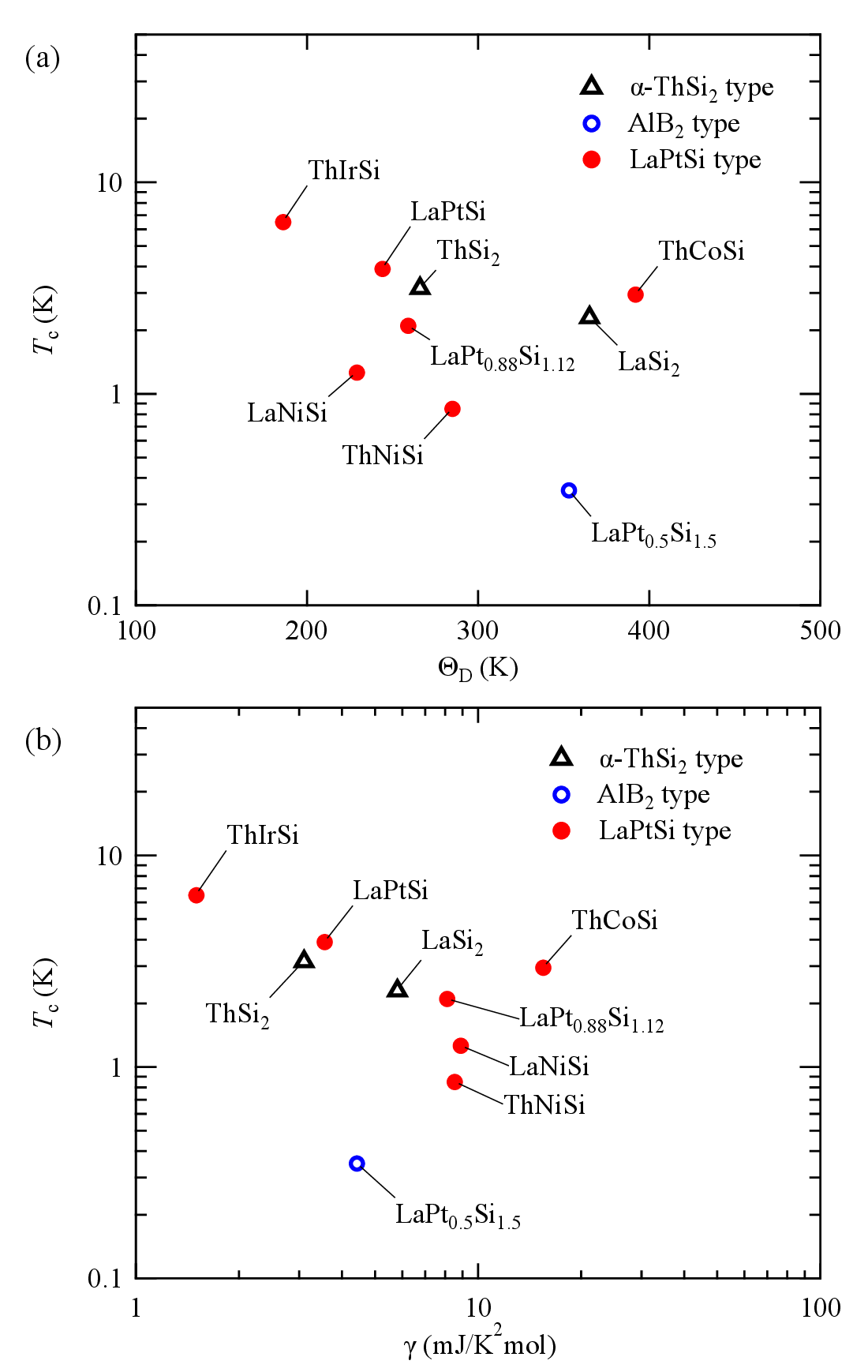}
 \caption{
(Color online) (a) Superconducting transition temperature $T_{\rm c}$ versus Debye temperature $\Theta_{\rm D}$, and (b) $T_{\rm c}$ versus the electronic specific heat coefficient $\gamma$ for 
noncentrosymmetric LaPtSi-type (tetragonal, $I4_{1}md$) compounds: 
LaPtSi,\cite{ramakrishnan1995a} ThIrSi,\cite{chevalier1986a} 
ThCoSi,\cite{Domieracki2016a} 
ThNiSi,\cite{Domieracki2018a} 
LaNiSi,\cite{lee1994a}
and LaPt$_{0.88}$Si$_{1.12}$ (present work); 
centrosymmetric $\alpha$-ThSi$_2$-type (tetragonal, $I4_{1}/amd$) compounds: 
LaSi$_2$ \cite{Ohtsuka1966a} and ThSi$_2$, \cite{chevalier1986a} 
and AlB$_2$-type (hexagonal, $P6/mmm$) LaPt$_{0.5}$Si$_{1.5}$ (present work). 
}
\label{fig:Tc_vs_gamma}
\end{figure}

Figure \ref{fig:Tc_vs_gamma} summarizes the dependence of $T_{\rm c}$ on $\Theta_{\rm D}$ and $\gamma$ for various compounds, including tetragonal LaPtSi-type compounds, \cite{ramakrishnan1995a,chevalier1986a,Domieracki2016a,Domieracki2018a,lee1994a}
tetragonal $\alpha$-ThSi$_2$-type compounds, \cite{Ohtsuka1966a,chevalier1986a}
and hexagonal AlB$_2$-type LaPt$_{0.5}$Si$_{1.5}$.
From Fig. \ref{fig:Tc_vs_gamma}(a), it can be seen that lower values of $\Theta_{\rm D}$ tend to correspond to higher values of $T_{\rm c}$. 
ThIrSi exhibits the highest $T_{\rm c}$ and the lowest $\Theta_{\rm D}$ among the LaPtSi-type compounds. 
This trend is also evident in the present LaPt$_{x}$Si$_{2-x}$ system: for LaPtSi and LaPt$_{0.88}$Si$_{1.12}$, a lower $\Theta_{\rm D}$ corresponds to a higher $T_{\rm c}$. 
Such a relationship between $T_{\rm c}$ and $\Theta_{\rm D}$ has also been observed in other systems, such as BaNi$_2$(As$_{1-x}$P$_{x}$)$_{2}$ \cite{Kudo2012a} and AEPd$_{2}$As$_{2}$ (AE = Ba, Sr, and Ca). \cite{Kudo2017a}
In these systems, a lower $\Theta_{\rm D}$, which corresponds to low-lying phonons, results in a higher $T_{\rm c}$ due to strong electron-phonon coupling. 
The signature of strong coupling is evident in ThIrSi, as indicated by the enhanced specific heat jump $\Delta C / \gamma T_{\rm c}$ = 4.73,\cite{chevalier1986a} which is significantly larger than the BCS weak-coupling limit of 1.43. In contrast, the specific heat jump for the present LaPt$_{0.88}$Si$_{1.12}$ is close to the BCS weak-coupling limit, as mentioned above. 
Figure \ref{fig:Tc_vs_gamma}(b) shows the plot of $T_{\rm c}$ versus $\gamma$, which reveals an unusual trend: smaller $\gamma$ values correspond to higher $T_{\rm c}$. This is contrary to the typical trend in phonon-mediated superconductors, where a higher electronic density of states (DOS) at the Fermi level $E_{\rm F}$, and thus a larger $\gamma$, is associated with a higher $T_{\rm c}$. \cite{takagi1997a}
This suggests that in the family of LaPtSi-type compounds, the key factor determining $T_{\rm c}$ is not the electronic DOS at $E_{\rm F}$, but rather the phonon properties, particularly the Debye temperature ($\Theta_{\rm D}$). This is consistent with theoretical calculations of the electron-phonon coupling constant and $T_{\rm c}$.\cite{zhang2020a}
Zhang et al. demonstrated that the strongest electron-phonon coupling in LaPtSi originates from a phonon mode associated with the Pt motion.\cite{zhang2020a}
Accordingly, the observed decrease in $T_{\rm c}$ with decreasing $x$ from 1.0 to 0.86 in tetragonal LaPt$_{x}$Si$_{2-x}$ can be attributed to changes in this phonon mode, which is likely modified due to the partial occupation of the Pt site in the ordered hyper-honeycomb network, shown in Fig. \ref{fig:structures}(c), by Si atoms.

\section{\label{sec:Laptsi_conclusions}Conclusions}

To summarize, we explored the crystal structures and superconductivity (SC) in the solid solutions of LaPt$_x$Si$_{2-x}$. 
We found that the site disorder (Pt/Si) in crystals of LaPt$_x$Si$_{2-x}$ significantly reduces the SC transition temperature in the non-centrosymmetric tetragonal phase, as compared to similar systems like ThRh$_x$Si$_{2-x}$ and ThIr$_x$Si$_{2-x}$.\cite{lejay1983a,chevalier1986a} 
Much like ThCo$_{x}$Si$_{2-x}$,\cite{zhong1985a} a hexagonal phase of an AlB$_{2}$-type structure exists in LaPt$_x$Si$_{2-x}$ at around $x = 0.5$ between the centrosymmetric and non-centrosymmetric tetragonal phases. 
As seen from resistivity and susceptibility measurements, the hexagonal phase is also a superconductor with its $T_{\rm c}$ around 0.38 K. 
Such a discovery of a hexagonal phase with an AlB$_2$-type structure within a small range of $x$ has also been reported for ThCo$_{x}$Si$_{2-x}$, with an onset of SC around 2 K.\cite{zhong1985a} 
SrSi$_2$ is another system that is a semiconductor in the cubic phase ($P4_{1}32$, $O^{7}$, \#213); however, the partial substitution of Ni for Si results in a hexagonal phase of an AlB$_{2}$-type structure, which is superconducting at 2.8 K.\cite{pyon2012a} Our observation of a new superconductor, LaPt$_{0.5}$Si$_{1.5}$, in the hexagonal phase is an interesting addition to these families.\\

\begin{acknowledgment}
SXRD data was collected at SAIF laboratory, IIT Madras, India. 
EPMA was performed at N-BARD, Hiroshima University, Japan (NBARD-00191).
This work was supported by JSPS KAKENHI (Grant Nos. JP23H04630, JP23H04861, JP23H04870, JP22H01168, JP22K03529, JP22H04473, and JP21K03448), 
JGC-S Scholarship Foundation (No. 2010),
The Hattori Hokokai Foundation (No. 21-010), and
The Mazda Foundation (No. 21KK-191).
\end{acknowledgment}

\appendix

\section{Synthesis}
\label{app:synthesis}
\begin{figure}
\centering
\includegraphics[width=80mm]{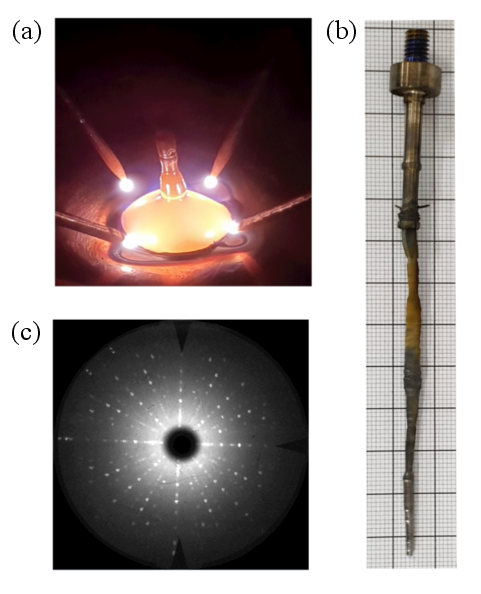}
 \caption{(Color online) (a) Crystal growth in a tetra arc furnace by the Czochralski method. 
 (b) A cylindrical shaped ingot was obtained of about 70 mm in length and 4 mm in diameter.
 (c) Laue pattern corresponding to (001) plane suggesting a high quality single crystal.}
\label{fig:crystalgrowth}
\end{figure}

Single crystal of LaPt$_{0.88}$Si$_{1.12}$ was synthesized by the Czochralski (Cz) method in a tetra-arc furnace (Techno Search Corporation, Japan) under ultra-pure Ar atmosphere, following an iterative process as depicted in Fig. \ref{fig:crystalgrowth} (a).
Initially, high-purity elements La:Pt:Si (99.99\% for La and Pt, and 99.999\% for Si) were taken in a stoichiometric ratio of 1:1:1, totaling 10 g, and melted repeatedly to ensure homogeneity.
A polycrystalline seed crystal was then cut from this ingot for crystal growth.
The polycrystalline seed was gently inserted into the molten solution and initially pulled at a rapid speed of about 90 mm/h. The melt temperature was adjusted to form a neck, and then the pulling speed was reduced to approximately 10 mm/h for the remainder of the growth process.
A 70 mm long ingot was pulled with a diameter of 3--4 mm, as shown in Fig.~\ref{fig:crystalgrowth}(b). 
Laue diffraction confirmed that the crystal exhibited good single crystallinity, as shown in Fig.~\ref{fig:crystalgrowth}(c).

Polycrystalline samples of LaPt$_{x}$Si$_{2-x}$ ($x$ = 0.50, 0.75, 0.80, 0.85, and 1.00) were synthesized by arc-melting in an ultra-pure Ar atmosphere.
Initially, high-purity elements La:Pt:Si (99.99\% for La, and 99.999\% for Pt and Si) were taken in ratios of 1:$x$:$2-x$.
The buttons were flipped and melted several times to ensure homogeneity.
Subsequently, they were wrapped in Ta foil within a quartz ampoule and annealed at 950°C for 1 week.
Cutting of both single crystals and polycrystals was performed using a wire electric discharge machine.

\section{Single-crystal X-ray diffraction}
\label{app:sxrd}

Small pieces of single crystals were acquired by crushing a large single crystal, from which a crystal of dimensions
0.080 $\times$  0.062 $\times$ 0.043 mm$^{3}$ was selected for single-crystal X-ray diffraction (SXRD) experiment
at room temperature.
SXRD was measured on a four-circle Bruker diffractometer employing Mo-K$\alpha$ (0.71073 \AA) radiation.
Diffracted X-rays were detected by a Photon II detector where the crystal-to-detector distance was 50 mm, resulting in a resolution of the SXRD data of approximately $(\sin(\theta)/\lambda)_{\rm max}$ = 0.694290 \AA$^{-1}$. 
See Table \ref{stab:laptsi_crystalinfo} for the crystallographic information.

\begin{table}
\caption{
Crystallographic data of
LaPt$_{0.88}$Si$_{1.12}$.}
\label{stab:laptsi_crystalinfo}%
\begin{tabular}{cccc}
\hline \hline
Temperature (K)  & 300 & &\\
Crystal system & Tetragonal& & \\
Space group & $I4{_1}md$ & &\\
Space group No. & 109 & &\\
$a$ (\AA{})  &4.2441(2) & & \\
$c$ (\AA{})  &14.5264(2)  & &  \\
Volume (\AA{}$^3$)   & 261.66 (2) &&  \\
$Z$ & 4  & & \\
Wavelength (\AA{}) & 0.71073  & & \\
Detector distance (mm) &50 & & \\
$\theta$-range (deg) &5.004 to 29.568 & & \\
Rotation per image (deg) & 0.5  & &\\
$(\sin(\theta)/\lambda)_{\rm max}$ (\AA{}$^{-1}$) & 0.694290  & &\\
Absorption, $\mu$ (mm$^{-1}$)  & 63.588 &  & \\
T$_{\rm min}$, T$_{\rm max}$ &  0.032, 0.102 & & \\
Criterion of observability & $I>3\sigma(I)$ & &\\
Number of reflections \\
measured  & 2777 & \\
unique (obs/all)  &103/119 & & \\
$R_{\rm int}$ (obs/all)  &0.0620/0.0629 & &\\
No. of parameters &15  & & \\
$R_{F }$  (obs)  &0.0311 & &  \\
$wR_{F }$ (all)  &0.0509 &  &\\
GoF (obs/all)& 3.67/3.49  &  & \\
$\Delta\rho_{\rm min}$, $\Delta\rho_{\rm max}$(e \AA$^{-3}$)  &-3.95, 5.86 & &  \\
\hline \hline
\end{tabular}
\end{table}

\section{Electron-probe micro-analysis (EPMA)}
\label{app:epma}

\begin{figure}
\centering
\includegraphics[width=100mm]{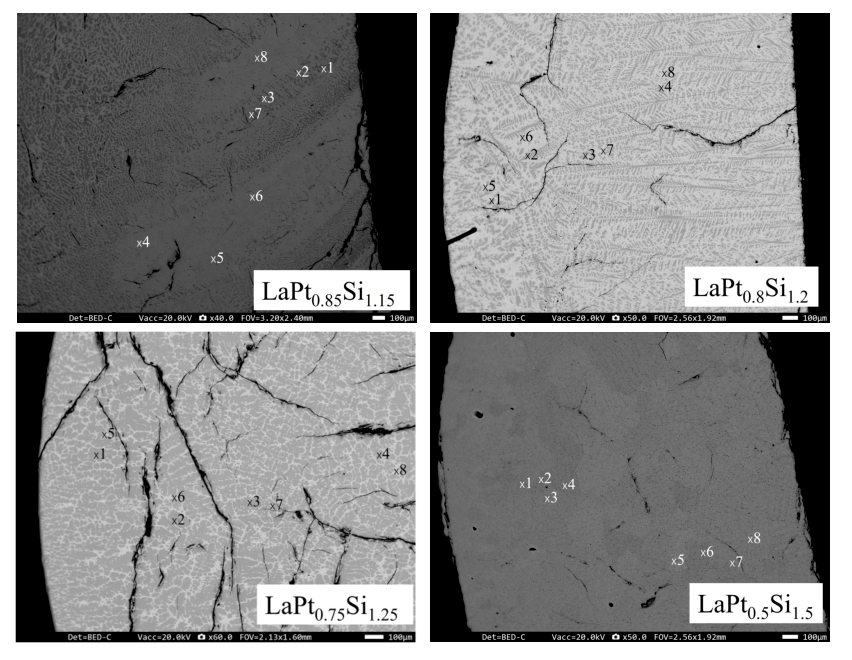}
\caption{
Back scattered electron images of LaPt$_{x}$Si$_{2-x}$ with $x_{\rm nominal}$ = 0.50, 0.75, 0.80, and 0.85. 
}
\label{fig:epma}
\end{figure}

Figure \ref{fig:epma} shows back scattered electron images of $x_{\rm nominal}$ = 0.50, 0.75, 0.80, 0.85. 
Points are labelled in the figure and analyzed. 
Table \ref{table:epma} shows the results of the analysis.
One observes a clear phase seperation for $x_{\rm nominal}$ = 0.75, 0.80, and 0.85, while for $x_{\rm nominal}$ = 0.5 the phase is singular phase with minimal difference between $x_{\rm nominal}$ and $x_{\rm analyzed}$.

\begin{table}
\caption{
Local composition $x$ in LaPt$_{x}$Si$_{2-x}$ analyzed by EPMA for $x_{\rm nominal}$ = 0.85, 0.80, 0.75, and 0.50. 
Points measured are indicated in Fig. \ref{fig:epma}.}
\label{table:epma}%
\begin{tabular}{ccccc}
\hline
Point no.&  LaPt$_{0.85}$Si$_{1.15}$ &  LaPt$_{0.80}$Si$_{1.20}$ 
&  LaPt$_{0.75}$Si$_{1.25}$  &  LaPt$_{0.50}$Si$_{1.50}$  \\
\hline
x1 & 0.709    & 0.694    & 0.697      & 0.499  \\
x2 & 0.703    & 0.687    & 0.691      & 0.493  \\
x3 & 0.709    & 0.697    & 0.695      & 0.492  \\
x4 & 0.828    & 0.692    & 0.690      & 0.482  \\
x5 & 0.821    & 0.859    & 0.870      & 0.478  \\
x6 & 0.828    & 0.871    & 0.858      & 0.498  \\
x7 & 0.820    & 0.867    & 0.873      & 0.501  \\
x8 & 0.831    & 0.871    & 0.871      & 0.504  \\
\hline
\end{tabular}
\end{table}

\section{\label{app:pxrd}Powder X-ray diffraction}
\label{app:pxrd}

\begin{figure}
\centering
\includegraphics[width=90mm]{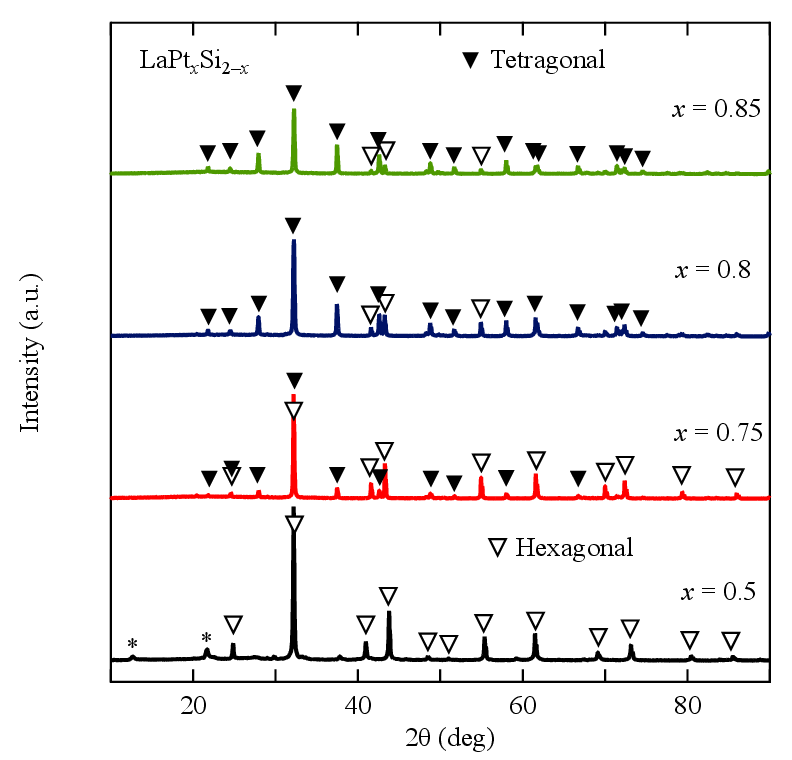}
 \caption{(Color online) 
PXRD pattern of LaPt$_{x}$Si$_{2-x}$ for $x_{\rm nominal}$ = 0.50, 0.75, 0.80, and 0.85. 
The filled and open triangles indicate the tetragonal and hexagonal peaks, respectively.
Phase separation can be observed for all except for $x$ = 0.5 where it is in the pure hexagonal phase. 
}
\label{fig:PXRD}
\end{figure}

Figure \ref{fig:PXRD}  shows PXRD pattern of LaPt$_{x}$Si$_{2-x}$ with $x_{\rm nominal}$ = 0.50, 0.75, 0.80, and 0.85. 
The diffraction pattern for $x_{\rm nominal}$ = 0.50 demonstrates the pure hexagonal phase, while the patterns for $x_{\rm nominal}$ = 0.75, 0.80, and 0.85 exhibit both hexagonal and tetragonal diffractions. The peak intensities of hexagonal phase  diminish with increasing Pt content $x$. 

\end{document}